\documentclass[aps,prd,twocolumn,amsfonts,amssymb,amsmath,showpacs,letterpaper,superscriptaddress,nofootinbib,altaffilletter,
fontenc]{revtex4-1}

\usepackage{amsmath,amssymb,graphicx}
\usepackage[english]{babel} % English language/hyphenation
\usepackage[printwatermark]{xwatermark}
\usepackage{xcolor}
\usepackage{lipsum}
\usepackage{bm}
\usepackage{multirow}
\usepackage{color}
\usepackage{booktabs}
\usepackage{tabularx}
\usepackage{times}
\usepackage{microtype}
\usepackage{booktabs}
\usepackage{subfigure}
\usepackage[normalem]{ulem}
\usepackage[varg]{txfonts}
\usepackage{dirtytalk}
\usepackage[colorlinks, pdfborder={0 0 0}]{hyperref}
\definecolor{LinkColor}{rgb}{0.75 , 0, 0}
\definecolor{CiteColor}{rgb}{0, 0.5, 0.5}
\definecolor{UrlColor}{rgb}{0, 0, 0.75}
\hypersetup{linkcolor=LinkColor}
\hypersetup{citecolor=CiteColor}
\hypersetup{urlcolor=UrlColor}
\usepackage[utf8]{inputenc}
\usepackage{csquotes}
\usepackage[normalem]{ulem}

\colorlet{Mycolor1}{green!10!orange!90!}
\definecolor{Mycolor2}{HTML}{00F9DE}

\allowdisplaybreaks
\newcommand{\FigStart}{\begin{figure*}[h]}
\newcommand{\FigEnd}{\end{figure*}}

\def\dks{\delta\kappa_s}
\def\dka{\delta\kappa_a}
\def\dkOne{\delta\kappa_1}
\def\dkTwo{\delta\kappa_2}
\def\dkOneTwo{\delta\kappa_{1,2}}

\def\imrPhenomPv2{\texttt{IMRPhenomPv2}}
\def\taylorF2{\texttt{TaylorF2}}

\def\dj{d^{(j)}}

\def\barAlpha{\vec{\alpha}}
\def\dataN{\vec{d}}
\def\dj{d_{\rm j}}
\def\hj{{h}_{\rm j}}

\def\fnbh{f_{\text{nbh}}}
\def\pbh{\vec{\theta}}
\def\pnbh{\pbh,\dks}
\def\Znbh{\mathcal{Z}^{nbh}}
\def\Zbh{\mathcal{Z}^{bh}}
\def\Zjnbh{\Znbh_{\rm j}}
\def\Zjbh{\Zbh_{\rm j}}
\def\Uniform{\texttt{Uniform}}
\def\GausPos{\texttt{GausPos}}
\def\GausNeg{\texttt{GausNeg}}

\def\BBH{\texttt{BBH}}
\def\NonBBH{\texttt{NonBBH}}
\def\NonBBHPos{\texttt{NonBBHPos}}
\def\NonBBHNeg{\texttt{NonBBHNeg}}
\def\MixtureAll{\texttt{MixtureAll}}
\def\MixturePos{\texttt{MixturePos}}

\begin{document}
\title{
%or\\
%Inferring the nature of a compact binary population using spin-induced quadrupole moment measurements
{Population inference of spin-induced quadrupole moments as a probe for non-black hole compact binaries}} 
\author{Muhammed Saleem}\email{mcholayi@umn.edu}
\affiliation{School of Physics and Astronomy, University of Minnesota, Minneapolis, MN 55455, USA}
\affiliation{Chennai Mathematical Institute, Siruseri 603103, Tamilnadu,
India}
\author{N. V. Krishnendu}\email{krnava@aei.mpg.de}
\affiliation{Max Planck Institute for Gravitational Physics (Albert Einstein Institute), Callinstr. 38, 30167 Hannover, Germany}
\affiliation{Leibniz Universitat Hannover, D-30167 Hannover, Germany}
\author{Abhirup Ghosh}
\affiliation{Max Planck Institute for Gravitational Physics (Albert Einstein Institute),  Am Mühlenberg 1, 14476 Potsdam, Germany}
\author{Anuradha Gupta}
\affiliation{Department of Physics and Astronomy, The University of Mississippi, Oxford MS 38677, USA}
\author{W. Del Pozzo}
\affiliation{Dipartimento di Fisica “Enrico Fermi”, Universita` di Pisa, and INFN sezione di Pisa, Pisa I-56127, Italy}
\affiliation{Institute of Gravitational Wave Astronomy, University of Birmingham, Edgbaston, Birmingham B15 2TT, United Kingdom}
%\author{Archisman Ghosh}
%\affiliation{Nikhef, Amsterdam}
\author{Archisman Ghosh}
\affiliation{Ghent University, Proeftuinstraat 86, 9000 Gent, Belgium}
%\affiliation{Lorentz Institute, Leiden University, PO Box 9506, Leiden 2300 RA, The Netherlands}
%\affiliation{GRAPPA, University of Amsterdam, Science Park 904, 1090 GL Amsterdam, The Netherlands}
%\affiliation{Delta Institute for Theoretical Physics, Science Park 904, 1090 GL Amsterdam, The Netherlands}
%\affiliation{Nikhef – National Institute for Subatomic Physics, Science Park 105, 1098 XG Amsterdam, The Netherlands}
\author{K. G. Arun}\email{kgarun@cmi.ac.in}
\affiliation{Chennai Mathematical Institute, Siruseri 603103, Tamilnadu, India}

\date{\today}

%%%%%%%%%%%%%%%%%%%%%%%%%%%%%%%%%%%%%%%%%%%%%%%%%%%%%%%%%%%%%%%%%%%%%%%%%%%%%
\begin{abstract}
 Gravitational wave (GW) measurements of physical effects such as spin-induced quadrupole moments can distinguish binaries consisting of black holes from non-black hole binaries.
    While these effects may be poorly constrained for single-event inferences with the second-generation detectors, combining information from multiple detections can help uncover features of non-black hole binaries. The spin-induced quadrupole moment has specific predictions for different types of compact objects, and a generalized formalism must consider a population where different types of compact objects co-exist. 
    In this study, we introduce a hierarchical mixture-likelihood formalism to estimate the {\it fraction of non- binary black holes in the population}. We demonstrate the applicability of this method using simulated GW {signals} injected into Gaussian noise following the design sensitivities of the Advanced LIGO Advanced Virgo detectors. We compare the performance of this method with a traditionally-followed hierarchical inference approach. Both the methods are equally effective to hint at  inhomogeneous populations, however, we find the mixture-likelihood approach to be more  natural for mixture populations comprising compact objects of diverse classes. 
    We also discuss the possible systematics in the mixture-likelihood approach, caused by several reasons, including the limited sensitivity of the second-generation detectors, specific features of the astrophysical population distributions, and the limitations posed by the  waveform models  employed. Finally, we apply this method to the LIGO-Virgo detections published in the second GW transient catalog (GWTC-2) and find them consistent with a binary black hole population within the statistical precision.

\end{abstract}

 \maketitle
%%%%%%%%%%%%%%%%%%%%%%%%%%%%%%%%%%%%%%%%%%%%%%%%%%%%%%%%%%%%%%%%%%%%%%%%%%%%%%%%%
\section{Introduction}\label{intro}

Gravitational-wave (GW) observations are slated to unravel a plethora of compact binaries in the coming years. The LIGO-Virgo-KAGRA Collaboration
has already observed several binary black holes (BBHs) \cite{Discovery,GW151226,LIGOScientific:2020ibl,GW170104,GW170608,GW170814,O1O2catalogLSC2018,GW190412,GW190814, GW190521}, binary neutron stars~\cite{GW170817,GW190425}  and neutron star - black hole mergers \cite{LIGOScientific:2021qlt}.
However, one may wonder if there are compact objects other than black holes (BHs) and neutron stars (NSs) that are made of exotic matter or described by some unknown physics\footnote{White dwarfs, whose sizes are larger than their gravitational radii, are not considered as compact objects in current gravitational wave astronomy.}.  There are theoretical predictions of exotic compact objects which can mimic properties of BHs and are referred as BH mimickers~\cite{CardosoLivingReview,Giudice} (\textit{a.k.a} non-BH compact objects). Some examples include boson stars~\cite{Jetzer:1991jr,Liddle:1993ha,Schunck:2003kk,Liddle:1993ha,Liebling:2012fv,Brito:2015pxa,Friedberg:1986tp,Friedberg:1986tq,Ryan97b,Visinelli:2014twa}, fermionic stars~\cite{Bauswein:2009im,GondekRosinska:2008nf,Weber:2004kj,Madsen:1998uh,Alcock:1986hz,Alford:2001dt},  multi-component stars~\cite{Kouvaris:2015rea,Freese:2015mta}, dark energy stars~\cite{Chirenti:2016hzd,Chirenti:2007mk,Lobo:2005uf,Carter:2005pi,Visser:2003ge,Mazur:2001fv} and dark matter stars~\cite{Kouvaris:2015rea,Tulin:2013teo,Bertone:2019irm,Barack:2018yly,Ferrer:2017xwm,Spolyar:2007qv}. The unknown intrinsic properties of these objects are expected to be imprinted in the GWs they emit and hence GW observations of such objects provide a unique way to probe their presence~\cite{Krishnendu:2017shb, Cardoso:2017cfl,Sennett:2017etc,Johnson-McDaniel:2018uvs}.

 The compact binary mergers observed by the advanced LIGO~\cite{Aasi:2013wya,2015} and the advanced Virgo detectors~\cite{TheVirgo:2014hva} in the first three observing runs are all consistent with mergers composed of BHs and NSs~\cite{Discovery,GW151226,GW170104,GW170608,GW170814,GW170817,O1O2catalogLSC2018,GW190425,GW190412,GW190814, GW190521,IAScatalog,LIGOScientific:2020ibl}.  The current sensitivity of detectors is insufficient to rule out the presence of signals from exotic compact objects in the data.  With more of such detections in the future~\cite{Aasi:2013wya,Sathyaprakash:2011bh,Regimbau:2012ir, Hild:2010id, Hild:2008ng,Evans:2016mbw,DecidoPathFinder2009,LPF2017,LPFFirstResultsPRL,LISApathfinder2019March}, one of the important science goals would be to look for the existence of exotic compact objects in the data~\cite{Krishnendu:2019tjp}. These observations may allow us to constrain what fraction of the detected events could be exotic compact objects. In turn, this can shed light on some of the unexplored physics realms concerning exotic particles and dark matter physics~\cite{Ryan:1995wh,Collins:2004ex,Vigeland:2009pr,Glampedakis:2005cf,Cardoso:2017cfl,Sennett:2017etc,Jimenez-Forteza:2018buh,Abdelsalhin:2018reg,Johnson-McDaniel:2018uvs,Porto:2016zng,Chirenti:2007mk,BertiBS,Macedo:2013jja,Macedo:2016wgh,BertiBS,GossanVeitchSathya2011ha,RelativistictheoryofTLNs_EricBinnington,NHTBHsAstrophysicalEnvironments2015,DamourNagarTLNsofNSs,Hinderer:2007mb,FlanaganHindererNSTLNs,Pani:2015hfa,Hartle:1973zz,Maselli:2017cmm,Chatziioannou:2016kem,Chatziioannou:2012gq,Laarakkers:1997hb,Pappas:2012qg,Pappas:2012ns,Narikawa:2021pak}.

There have been remarkable progress in the modeling of BBH waveforms in the inspiral~\cite{Bliving,Marsat:2012fn,Bohe:2012mr,Bohe:2013cla,Marsat:2013caa,Bohe:2015ana,Marsat:2014xea,Arun:2008kb,Kidder:1995zr,Will:1996zj,Buonanno:2012rv,Blanchet:2013haa,Mishra:2016whh,Buonanno:1998gg,Nagar:2018zoe,EOB_Precessing_2013,Blanchet}, merger~\cite{Pretorius:2005gq}  and ringdown~\cite{TSLivRev03,Sasaki:2003xr} regimes in general relativity (GR). There has also been much progress in modeling binaries containing neutron stars~\cite{Dietrich:2018uni,PhenomNSBH,Dietrich:2017aum}. But the same is not true for binaries of BH mimickers, where the progress has been slow, primarily due to the mathematical complications these objects pose in the modeling~\cite{Cardoso:2019rvt,Pacilio:2020jza,Sennett:2017etc,Balakrishna:2006ru,Abedi:2016hgu}. Therefore, looking for the exact imprints of the BH mimicker models in the observed GW signals is difficult.

In the post-Newtonian (PN) formalism, the effects distinguishing BHs from BH mimickers are well studied. These effects include the deformations of the compact objects due to the tidal field of the companion~\cite{Cardoso:2017cfl, Johnson-McDaniel:2018uvs}, or its spinning motion and the effects of tidal heating~\cite{Hartle:1973zz,Maselli:2017cmm,Chatziioannou:2016kem,Chatziioannou:2012gq}. A number of {\it tests} have been proposed to distinguish BH mimickers from BHs using parametrizations of such  physical effects. Examples include the tests based on the tidal deformability measurements~\cite{RelativistictheoryofTLNs_EricBinnington,NHTBHsAstrophysicalEnvironments2015,DamourNagarTLNsofNSs,Hinderer:2007mb,FlanaganHindererNSTLNs,Pani:2015hfa, Cardoso:2017cfl,Sennett:2017etc,Jimenez-Forteza:2018buh,Abdelsalhin:2018reg,Johnson-McDaniel:2018uvs,Porto:2016zng,GW170817,GW170817LIGOScientificModelSel,Abdelsalhin:2019ryu,Huang:2020cab,Hinderer:2007mb,FlanaganHindererNSTLNs,Pani:2015hfa, Cardoso:2017cfl,Jimenez-Forteza:2018buh,Abdelsalhin:2018reg}, tidal heating parameter estimations~\cite{Hartle:1973zz,Maselli:2017cmm,Chatziioannou:2016kem,Chatziioannou:2012gq,Datta:2020gem}, etc. There are also methods based on the inference of the late-ringdown echo parameters~\cite{Cardoso:2017cqb}. 
 
In this work, we follow a method that uses spin-induced quadrupole moment parameter to distinguish BBHs from binaries of BH mimickers, as outlined in \cite{Krishnendu:2017shb, Krishnendu:2019ebd}. The spin-induced quadrupole moment parameter has a unique value, unity, for Kerr BH according to the no-hair conjecture
\cite{Hansen74,Carter71,Gurlebeck:2015xpa}, whereas, for any other compact object, its value can be different from unity. A Bayesian framework to measure this parameter has been comprehensively demonstrated in a previous study~\cite{Krishnendu:2019tjp} using simulated GW signals. {Moreover, the constraints on the spin-induced quadrupole moment parameters from the GWTC-2 events are reported in~\cite{Abbott:2020jks}}. In this work, our focus is on methods to combine their measurements from \emph{multiple} GW detections, which would be key in enhancing statistical evidence \emph{in favor of} or \emph{against} BH mimickers. In particular, the fact that BHs and BH mimickers can co-exist in the universe, points to the need for having a generic framework that can unravel how various compact objects are distributed in the universe.

We discuss two methods to combine measurements from multiple detections. The first is a so-called  \emph{hierarchical combining approach} in which the spin-induced quadrupole moment parameters of the detected population are assumed to follow a Gaussian distribution and use a hierarchical framework to infer the moments of the distribution. This approach is  similar to the one proposed in~\cite{Isi:2019asy} and  has been used in \cite{LIGOScientific:2020tif} to infer the distribution of spin-induced quadrupole moment parameters for the LIGO-Virgo detected events.  
The second method, namely the \emph{mixture-likelihood approach},
explicitly assumes the population to be a mixture of  BBH and non-BBH mergers by parametrizing the fraction of events in the respective categories. Specifically, we use a parameter $\fnbh$ to quantify the fraction of non-BBH mergers in the detected population. Note that the mixture-likelihood approach is also hierarchical in nature, however, for the sake of name distinction, the notion of \emph{hierarchical} is henceforth used only for the former method.  

\begin{figure*}
    \centering
    \includegraphics[scale=0.6]{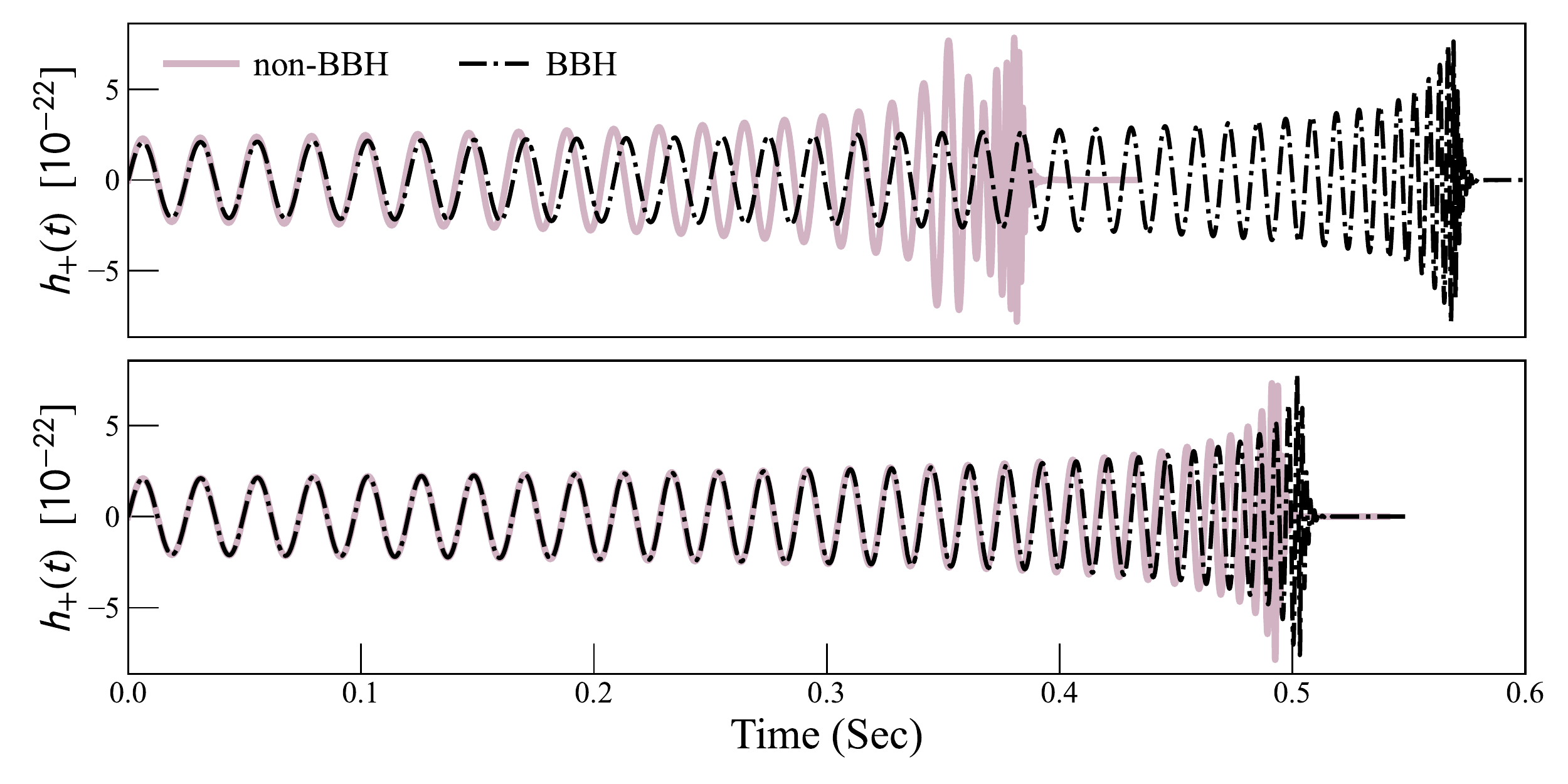}
    \caption{Time-domain gravitational waveforms for a fast-spinning (top) and slowly-spinning (bottom) compact binary mergers with component masses $(m_1, m_2) = (20, 10)M_{\odot}$. The spins are aligned with the orbital angular momentum vectors (no precession modulations) and have the dimensionless spin magnitudes $(\chi_1, \chi_2) = (0.6, 0.5)$  for the top  and $(\chi_1, \chi_2) =(0.15, 0.1)$ for the bottom panels. The black (dash-dot) traces are BBH waveforms, and the red (solid) traces are non-BBH waveforms with spin-induced quadrupole moment parameters assuming $\delta\kappa_1=40$ and $\delta \kappa_2=25$ for component compact objects. The time-domain waveforms are generated using backward fast Fourier transform (FFT) of the frequency-domain waveform model, \texttt{IMRPhenomPv2}. For each waveform, the time $t$ (x axis) is set to zero at a point when the instantaneous frequency of the waveform is 40 Hz, and the waveforms are also aligned to be in phase at that point.}
    
\label{fig:wf_plot}
\end{figure*} 

The  applicability of the two approaches are demonstrated using simulated  GW signals from binaries of compact objects of diverse classes. The hyperparameters that are used in both the methods can effeciently signal at the non-BH sub-populations that are present in the detected population. 
Furthermore, the \emph{mixture-likelihood approach} captures the complexity in the population powerfully and can capture the fraction of events that are from non-BH sub-populations.  However, some systematics are noticed with the \emph{mixture-likelihood approach}, which could be attributed to the limitations posed by the current detector sensitivities and some of the intrinsic properties of the astrophysical population.

The rest of this paper is organized as follows. We review the GW measurements of spin-induced quadrupole moment parameters in Sec.~\ref{sec:siqm}. Sec.~\ref{sec:method} describes the statistical methods employed in this study, including 
%the basics of Bayesian inference, 
the hierarchical and the mixture-likelihood approaches. 
In Sec.~\ref{sec:pop}, we detail the properties of simulated population of compact binary mergers used to for this analysis. We discuss the main findings in Sec.~\ref{sec:results} and discuss the systematic effects in Sec.~\ref{sec:bias}. We conclude the study in Sec.~\ref{sec:conc}
with a discussion on the future aspects. We also provide an appendix on how the mixture-likelihood approach would benefit by including astrophysical models of spin-induced quadrupole moments into the framework.

\section{Review: Gravitational-wave measurements of spin-induced multipole moments}\label{sec:siqm}

    The spin-induced multipole moments arise due to the  spinning motion of the compact objects in the binary and these effects appear in the gravitational waveform along with the self-spin terms. The leading order effect~\cite{Poisson:1997ha,Krishnendu:2017shb,Marsat:2014xea} at the second post-Newtonian (2PN) order can be schematically represented in the following form,
	\begin{equation} 
	    Q=-\kappa\,\mathbf{\chi}^2 \,m^3,
	\label{eq-leadingorder} 
	\end{equation} 
	where $Q$ is the spin-induced quadrupole moment scalar,
	$m$ is the mass and $\chi$ is the dimensionless spin  parameter, defined as $\vec{\chi}=\vec{S}/m^{2}$  where $\vec{S}$ is the spin angular momentum of the compact object.

    The spin-induced quadrupole moment parameter $\kappa$ has a unique value, unity, for Kerr BH according to the  no-hair conjecture
	\cite{Hansen74,Carter71,Gurlebeck:2015xpa}, whereas, for any other compact object, its value can be different from unity. For example, for spinning NSs, the value of $\kappa$ varies between $\sim 2-14$ depending upon the internal structure of the star~\cite{Laarakkers:1997hb,Pappas:2012qg,Pappas:2012ns}. Also, calculations show that the value of $\kappa$ can vary roughly between 10 to 150 for boson stars~\cite{Ryan97b} while for  gravastars~\cite{gravastars} $\kappa$ can be negative as well~\cite{Uchikata:2015yma, Uchikata2016qku}. In Fig.~\ref{fig:wf_plot} we compare gravitational waveforms of BBHs (i.e., $\kappa=1$) and non-BBHs ($\kappa\neq1$) for two different values of spin parameters. 
	%and shows how the gravitational wave signal changes with spins. 
 Both the binaries have same masses $(m_1, m_2) = (20, 10)M_{\odot}$ while non-BBHs have $(\kappa_1, \kappa_2) = (40, 25)$. Highly spinning binaries have $(\chi_1, \chi_2) = (0.6, 0.5)$ while slowly spinning binaries have $(\chi_1, \chi_2) =(0.15, 0.1)$, assuming the spins aligned to the orbital angular momentum axis. We used \texttt{IMRPhenomPv2} \cite{Ajith:2009bn,Husa:2015iqa,Hannam:2013oca} waveform model to simulate the time-domain GW signal. We see that as the spins of the binary components increase, the de-phasing between the BBH and non-BBH waveforms increases.

	A Bayesian framework to measure the $\kappa$ parameters was demonstrated in ~\cite{Krishnendu:2019tjp}, to constrain the nature of the stellar mass compact binaries detected by Advanced LIGO and Advanced Virgo detectors. It was shown that the spin-induced quadrupole moment measurements can be used to distinguish the observed BBHs from non-BBHs for inspiral dominated systems with moderate to high spins \cite{Krishnendu:2019tjp}.
	
    In this framework, one uses BBH waveforms which allow the spin-induced quadrupole moment coefficient $\kappa$ to vary around the expected Kerr value as $\kappa = 1+\delta\kappa$. Here the parametrized deformations (labeled as $\delta\kappa$) represents the deviations from the BH nature.  It is pointed out in \cite{Krishnendu:2019tjp} that the simultaneous measurement of $\delta\kappa_{1}$ and $\delta\kappa_{2}$ is difficult due to strong correlations between binary parameters in the gravitational waveform. In order to capture the deviation from the BBH nature Ref.~\cite{Krishnendu:2019tjp} proposed to measure their symmetric combination $\delta \kappa_s = (\delta\kappa_{1} + \delta\kappa_{2})/2$, assuming the anti-symmetric combination $\delta \kappa_a = (\delta\kappa_{1} - \delta\kappa_{2})/2$ vanishes for a BBH signal.  The $\delta \kappa_a = 0$ assumption also implies that the individual compact objects in the binary system are of the same nature. For cases with any violation of this assumption, we would expect an offset in $\dks$ posterior distribution, and such cases need more investigations keeping $\delta\kappa_{1}$ and $\delta\kappa_{2}$ as separate parameters. (See \cite{Krishnendu:2019ebd} for a more detailed discussion). The applicability of this test has been further explored in the context of expected detections from the third generation GW detectors~\cite{Krishnendu:2018nqa}. 

\section{Formalism}
\label{sec:method}

In a universe where all the massive compact objects are BBHs, $\delta\kappa_s$ assumes the unique and universal value $\delta\kappa_{s} = 0$ . However, if we admit the possibility that compact objects come in many flavors, this universality assumption would be wrong, and the inference on $\dks$ obtained by naively multiplying each observation's likelihood would lead to erroneous conclusions. Even for the binaries that are made up of  some specific class of exotic compact objects with a unique equation of state, the value of $\dks$ could vary depending on the intrinsic properties such as masses and spins as is found to be the case for boson stars~\cite{Ryan97b}. In such a case where the value of $\dks$ can vary from event to event, one would aim to infer the underlying distribution of  $\dks$ associated with the compact binary population. This is the context in which we discuss the two combining approaches and their applicability. 

Below, we introduce our notations for the Bayesian inference variables, followed by the formalisms for hierarchical combining and mixture-likelihood approach. 

\subsection{Bayesian inference: basic notations}\label{sec:method-basics}

    The first step in our formalism is to perform the Bayesian parameter estimation of all the detected GW events.  Here, we briefly overview the Bayesian inference method employed to estimate the spin-induced quadrupole moment parameter, $\dks$. 
    We define $\pbh$ as the vector representing the set of parameters that describes a BBH merger on quasi-circular orbits. This includes masses, spins, luminosity distance, time and phase of arrival, and the angles describing the sky-location and binary orientation. The data from the j$^{th}$ event is labeled as $\dj$, and the set of data from N events together is denoted as $\dataN$. $\mathcal{H}$ is our hypothesis that the data $\dj$  carries a signal $\hj(\pnbh)$ plus colored Gaussian random noise. 
    Under this hypothesis, the posterior for the binary parameters can be written as, 
    \begin{equation}
        p( \pnbh\, | \dj,\, \mathcal{H}) = \frac{\pi( \pnbh\, | \mathcal{H} ) \, \mathcal{L}(\dj\, |\, \pnbh, \, \mathcal{H}) }{\Zjnbh},
        \label{eq:posterior-nbh}
    \end{equation}
    where 
    $\mathcal{L}(\dj\, |\, \pnbh, \, \mathcal{H})$ is the likelihood of $\dj$ being the data given the parameters $\{\pnbh\}$, and $\pi( \pnbh\, | \mathcal{H} )$ is the prior probability of parameters \{$\vec \theta$. $\dks$\}.
    The evidence $\Zjnbh = P(\dj\, |\, \mathcal{H})$ is obtained by marginalizing the  likelihood over the prior, 
    \begin{equation}
        \Zjnbh = \int \pi(\pnbh) \, \mathcal{L}(\dj \vert \pnbh)\, d\pbh\, d\delta\kappa_s,  
        \label{eq:evidence}
    \end{equation}
    where, the superscript `\textit{nbh}' stands for the non-BBH  hypothesis  and we have dropped $\mathcal{H}$ for brevity. The BBH hypothesis is a special case obtained by fixing $\dks=0$ in the likelihood  $\mathcal{L}(\dj \vert \pbh, \delta\kappa_s=0)$ which we simply write as $\mathcal{L}(\dj \vert \pbh)$ and the corresponding evidence can be expressed as,
    \begin{equation}
        \Zjbh = \int \pi(\pbh) \, \mathcal{L}(\dj \vert \pbh)\, d\pbh \,.
        \label{eq:evidence-bh}
    \end{equation}
    The posterior on $\dks$ can be obtained by marginalizing Eq.~(\ref{eq:posterior-nbh}) over the BBH parameters as
    \begin{equation}
        p(\dks\, \vert \dj,\, \mathcal{H} ) = \int p( \pnbh\, | \dj,\, \mathcal{H})\, d\pbh.
        \label{eq:posterior-dks}
    \end{equation}

    {The \texttt{IMRPhenomPv2} signal model~\cite{Ajith:2009bn,Husa:2015iqa,Hannam:2013oca} used for the Bayesian analysis includes $\dks$ as a free parameter along with the BBH parameters $\pbh$. We assume a uniform prior on $\dks$ in $[-500, 500]$. For the component masses, we consider uniform priors on in the range $[4,100] M_{\odot}$. The priors on component spin-magnitudes are uniform in [0,1] and their orientations assumed to be isotropic. We choose uniform in co-moving volume ranging between $[10, 5000]$.}
    The parameter estimation is performed using the \verb|lalinference_nest| sampler available in the \texttt{LALInference} library package~\cite{Veitch:2014wba}. The posterior samples as well as the Bayesian evidences in Eq.~(\ref{eq:evidence}) and~(\ref{eq:evidence-bh}) are obtained as raw outputs from  \verb|lalinference_nest|. 

    Below, we discuss the two different approaches for combining $\dks$ measurements from the Bayesian analysis of individual events.

\subsection{Hierarchical combining approach: population distribution of $\delta\kappa_s$ }
\label{sec:method-hier}

In this approach, we assume that $\dks$ follows some underlying distribution governed by a set of hyper-parameters $\vec\alpha$, similar to the method demonstrated  in \cite{Isi:2019asy}.  
The posterior on $\barAlpha$ given the data $\dataN$ can be written as
\begin{equation}
p(  \barAlpha | \dataN ) \propto  \mathcal{L}(  \dataN | \barAlpha) \, p( \barAlpha),
\label{eq:post-baralpha}
\end{equation}
where the proportionality becomes equality by normalizing the right-hand side to unity. The prior $p( \barAlpha )$ is taken to be flat assuming no prior knowledge of the underlying distribution of the hyper-parameters $ \barAlpha$. Here $\mathcal{L}(  \dataN | \barAlpha)$ is the likelihood function which can be obtained as a product of likelihoods of $\barAlpha$ from individual events, as
\begin{equation}
\mathcal{L}( \dataN | \barAlpha) = \prod_{j=1}^N \mathcal{L}( \dj |  \barAlpha).
\label{eq:lhood-baralpha} 
\end{equation}
%\ag{shall we use the symbol $\cal L$ for likelihood?} 
which can be further expanded by re-writing the likelihood for the j$^{\text{th}}$ event as a marginalization over the $\dks$ parameter,
\begin{equation}
\mathcal{L}( \dataN | \barAlpha) = \prod_{j=1}^N 
\left[
\int \mathcal{L}(\dj | \dks )  \, p(\dks | \barAlpha ) \,d\dks
\right].
\label{eq:lhood-single}
\end{equation}
In the above equation, the term $\mathcal{L}(\dj | \dks )$ in the integral is the likelihood of $\dks$ for the j$^{th}$ event, marginalized over the BBH parameters. Since we use a uniform prior on $\dks$ in the single-event analyses, this likelihood will be the same as the posterior given in Eq.~(\ref{eq:posterior-dks}). The other term in the integral of   Eq.~(\ref{eq:lhood-single}), $p(\dks | \barAlpha )$ is the predicted distribution of $\dks$ given the hyper-parameters $\barAlpha$.  In this study, as mentioned earlier, we assume a Gaussian distribution with hyper-parameters $\barAlpha = \{\mu, \sigma\}$ which implies to,
\begin{equation}
    p(\dks | \barAlpha ) = \mathcal{N}(\mu, \sigma^2)\,.
    \label{eq:gaussian-single}
\end{equation}
In this study, we assume $\mu$ and $\sigma$ to have uniform priors in the ranges $[-150,150]$ and $[0,300]$  respectively and we use the \texttt{Dynesty} sampler from \texttt{Bilby} to sample over $\mu$ and $\sigma$ as per the likelihood given in Eq.~(\ref{eq:lhood-single}). 

Once the posterior of $\mu$ and $\sigma$ (or $\barAlpha$ in general) is computed, the population distribution of $\dks$ can be obtained as, 
\begin{equation}
    p( \dks | \dataN ) = \int p( \dks | \barAlpha )\, p(\barAlpha | \dataN ) d\barAlpha\,,
\label{eq:post-dks}
\end{equation}
where we have marginalized $\dks$ over the inferred distributions of the hyper-parameters $\barAlpha$. 

\subsection{Mixture-likelihood approach: Estimating the fraction of non-BBH events}
\label{sec:method-fraction}

Unlike the hierarchical approach in the preceding section, here we ask a more generic question ``what {\it fraction} of the detected population are from non-BBH events?''. We try to answer this with a \textit{mixture-likelihood} which parametrizes the presence of non-BBH events as $\fnbh$, defined as the {\it fraction} of total detected signals that are from non-BBH events. Mixture likelihoods have been used in literature for various problems, see e.g., Ref.~\cite{Smith:2017vfk}.  

Let's start with the single-event likelihood expression for the non-BBH model,
%\begin{widetext}
\begin{equation}
\mathcal{L}(\dj \vert \pbh, \delta\kappa_s) \propto \text{exp} \left({ -\frac{1}{2} \left(\dj-\hj(\pbh, \delta\kappa_s)\big\vert \dj-\hj(\pbh, \delta\kappa_s) \right) }\right).
\end{equation}
Here $( | )$ represents the noise-weighted inner product,  defined as
	$(x|y) = 4 \Re
	\int_{0}^{\infty} x(f)^*y(f)/\rm{S_{n}(f)} \, df 
$, where the $^*$ indicates complex conjugate and $\rm{S_{n}(f)}$ is the one-sided power spectral density (PSD) of the noise. %\ms{Not defined}. 
Suppose a fraction $\fnbh$ of the overall detectable signals are from non-BBH events, then the probability of any single event being a non-BBH will be equal to $\fnbh$. On the complementary side, the probability of any event being a BBH will be equal to  ($1-\fnbh$), as BBH and non-BBH are two mutually exclusive and exhaustive\footnote{Alternatively, one can consider specific non-BH models such as boson stars and/or gravastars. However, these models, together with the BH model, would still be non-exhaustive as yet unknown non-BH models could possibly exist.} cases.
 To take into account these possibilities, we can re-write the likelihood as a sum of the BBH and non-BBH likelihoods weighted by their respective probabilities as
%\begin{widetext}
\begin{equation}
\mathcal{L}(\dj \vert \pbh, \delta\kappa_s, \fnbh) = (1- \fnbh)\, \mathcal{L}(\dj \vert \pbh) + \fnbh \, \mathcal{L}(\dj \vert \pbh, \delta\kappa_s)\,.
\label{eq:mixture-single}
\end{equation}
%\end{widetext}
Equation~(\ref{eq:mixture-single}) is the single-event mixture-likelihood. 
%\ms{adding further equations (below) to explain the steps in between. Needed?}{\nvk {I think it is better to keep these extra equations}}
{Marginalizing over $\pbh$ and $\delta\kappa_s$, the above expression becomes,
\begin{eqnarray}
\mathcal{L}(\dj \vert \fnbh) &=& (1-\fnbh) \times \int \pi(\pbh) \, \mathcal{L}(\dj \vert \pbh)\, d\pbh \nonumber \\ 
&+& \fnbh \times \int \pi(\pnbh) \, \mathcal{L}(\dj \vert \pnbh)\, d\pbh\, d\delta\kappa_s, 
\label{eq:mixture-interm-1}
\end{eqnarray}
where the integrals on the R.H.S are the evidences for the BBH and non-BBH models, defined in  Eq.~(\ref{eq:evidence-bh}) and ~(\ref{eq:evidence}). Equation~(\ref{eq:mixture-single}) then becomes,} \begin{equation}
\mathcal{L}(\dj \vert \fnbh) = (1-\fnbh) \, \, \Zjbh + \fnbh \, \Zjnbh. \label{eq:mixture-interm-2}
\end{equation}
{Note that in going from Eq.~\ref{eq:mixture-interm-1} to Eq.~\ref{eq:mixture-interm-2}, we have used a uniform prior on $\dks$ as described in Section~\ref{sec:method-basics}. In Section~\ref{sec:bias}, we will investigate how the prior choices would affect the results.}

For a population of N detected events,  the combined likelihood can be written as, 
\begin{equation}
\mathcal{L}^{\text{pop}}(\dataN \vert \fnbh) = \prod_{j=1}^N \left( (1-\fnbh) \, \mathcal{Z}_j^{bh} + \fnbh \,\, \mathcal{Z}_j^{nbh} \right)\,.
\label{eq:likelihood-mixture}
\end{equation}
%\ag{maybe we could use $j$ as index instead of $i$ as $j$ was the event index }\ms{Correct!. Changed to j}
Equation~(\ref{eq:likelihood-mixture}) is the \textit{mixture-likelihood} for a population and can be evaluated for any value of $\fnbh$, by only knowing the evidences of BBH and non-BBH models for all the events in the population.  
With the above likelihood evaluated, we can express the posterior on $f_{nbh}$ as, \begin{equation}
p(\fnbh \vert \dataN) \propto \pi(\fnbh) \,\, \mathcal{L}^{\text{pop}}(\dataN \vert \fnbh)\,,
\label{eq:fnbh_post}
\end{equation}
where the prior on $\fnbh$ can be taken as uniform, $\mathcal{U}(0,1)$ owing to the most generic and uninformative case.

{Additionally, one can also define three mutually exclusive population-hypotheses based on the $\fnbh$ values. 
\begin{enumerate}
    \item $\fnbh = 0$: ``all the events are BBHs'',
    \item $0 < \fnbh < 1$: ``the population is a mixture of BBH and non-BBH events'',
    \item $\fnbh = 1$: ``all the events are non-BBHs''.
\end{enumerate}
It is straightforward to compute the Bayes factors between any of these two hypotheses.} The Bayes factor between 
``all are BBH'' and ``mixture of BBH and non-BBH events'' can be obtained as \begin{equation}\label{eq:logB-bh-vs-mix}
    \mathcal{B}_{mix}^{BBH} = \frac{ \mathcal{L}^{\text{pop}}(\dataN \vert \fnbh = 0) }  {\int\limits_{\tiny 0<\fnbh<1} \mathcal{L}^{\text{pop}}(\dataN \vert \fnbh) \, \pi(\fnbh) \, {\rm d}\fnbh }, 
\end{equation}
where the denominator included marginalizing the likelihood in Eq.~(\ref{eq:likelihood-mixture}) over the relevant range of $\fnbh$.
Similarly, the Bayes factor between ``all are BBH'' and ``all are non-BBH'' events can be obtained as, 
\begin{equation}\label{eq:logB-bh-vs-nbh}
    \mathcal{B}_{non-BBH}^{BBH} = \frac{ \mathcal{L}^{\text{pop}}(\dataN \vert \fnbh = 0) }{\mathcal{L}^{\text{pop}}(\dataN \vert \fnbh = 1) }\,.
\end{equation}

\section{Simulated compact binary population}
\label{sec:pop}

To demonstrate the performance of different combining approaches, we simulate a set of compact binary populations that include only BBH signals, only non-BBH signals, and different types of mixtures that include BBH and non-BBH signals at different proportions. In the subsections below, we describe the steps we followed to construct the populations. 

\subsection{Masses and spins}\label{sec:mass-spins}
\begin{enumerate}
\item 
We first choose a representative mass model from which we draw the component masses. The primary masses ($m_1$) follow  a distribution what is referred to as {\emph{Model-C}} in~\cite{o1o2_pop} which is a power-law function smoothened at the lower mass end and embedded with a Gaussian peak towards the higher mass end. The secondary masses ($m_2$) are drawn from a smoothened power-law conditional on the primary masses such that $m_1 \geq m_2$. 

\item 
For each binary, the magnitudes of the two component-spins are drawn according to the {\emph{Default Model }} as named in \cite{LIGOScientific:2020kqk}. In this model, the spin magnitudes  $a_{i}$ ($i={1,2}$)  are drawn from a beta distribution 
\begin{equation}
p(a_{i}|\alpha_{a},\beta_{a}) \propto a_{i}^{\alpha_{a}-1}(1-a_{i}^{\beta_{a}-1})\,, 
%B(\alpha_{a},\beta_{a}) \,,
\end{equation}
%(see Eq.~(D12) in~\cite{LIGOScientific:2020kqk}) 
where $\alpha_{a}$ and $\beta_{a}$ are shape parameters. 
%\ag{Is there a reason to have subscript a in $\alpha$ and $\beta$?}
We choose $\alpha_{a}=2.75$ and $\beta_{a}=6.00$  to make sure that we do not have sources with $a_{i}\sim 0$, as non-spinning compact objects do not carry imprints of spin-induced multipole moments.

\item 
The spin orientations are randomly drawn from a mixture of  {\it isotropic} and {\it aligned-to-orbital-angular-momentum}  orientations. In other words, the populations include binaries with precessing spins and binaries with non-precessing spins.

\end{enumerate}

Note that there are several mass and spin models in literature~\cite{LIGOScientific:2020kqk} which can explain the current GW data, and our choice here is arbitrary since they do not affect the conclusions of this study.

\subsection{ Source selection based on signal-to-noise-ratio}

The sources with masses and spins as described above are distributed uniformly in co-moving volume up to a redshift of 0.5~\cite{O1O2catalogLSC2018,LIGOScientific:2020kqk}.  The inclination and polarisation angles are chosen so that the binary orientations are isotropically distributed w.r.t detectors.

We construct our populations from sources that pass the following two criteria:
\begin{eqnarray}\label{eq:snr-cut}
	\rm{SNR} &\geq& 10, \nonumber  \\
   	\rm{SNR_{insp}} &\geq& 2 \,\rm{SNR_{post-insp}},
\end{eqnarray}
where, SNR is the optimal network signal-to-noise ratio with the HLV network, assuming all of them at their designed sensitivity \cite{Advanced_LIGO_Reference_Design, AdvancedLIGO, TheVirgostatus, TheVirgo:2014hva}. $\rm{SNR_{insp}}$  and  $\rm{SNR_{post-insp}}$ are the signal-to-noise ratio in the inspiral and post-inspiral  (merger-ringdown) regimes of the signal, respectively, determined by an inspiral cut-off frequency given by the inspiral to intermediate transition frequency of phenomenological waveform models. This cut-off frequency is calculated, given the total mass of the binary (M) as, $f_{\rm{cut}}=0.018/M$~\cite{Ajith:2009bn, Husa:2015iqa, Hannam:2013oca}.

The second criterion imposes the inspiral SNR to be at least twice the post-inspiral SNR, {which makes sure that there are enough number of waveform cycles in the inspiral phase.} This is because the spin-induced quadrupole moment effects predominantly affect the inspiral phase, as modeled in the current waveform models. However, with the advancements in numerical relativity simulations, future waveform models might accurately account for their evolution in the post-inspiral phase, which in turn would allow us to test {\it higher-mass} binaries whose SNR dominates in the post-inspiral phase.

\subsection{Distribution of spin-induced quadrupole moment parameters}

We simulate six random instances of the BBH populations with parameters as described in the preceding sections and apply the SNR criterion of Eq.~(\ref{eq:snr-cut}). We keep the first of these as a BBH population but turn the other five into either a non-BBH or a mixture population by associating spin-induced quadrupole moment parameters ($\dkOne$ and $\dkTwo$) to each of them. 
%\ag{do you mean $\kappa$? if yes, then there will be two parameters.}
Addition of these parameters would in principle change the SNR of the signals, however, the changes in our case are small\footnote{Due to the moderate to low spin values of the sources in our populations, we do not expect the power in the signals to change significantly, which is also apparent in the bottom panel of Fig.~\ref{fig:wf_plot}.} ($\leq$3\%) and all the signals still survive the SNR criterion.

The $\dkOne$ and $\dkTwo$ distributions for every populations are made by mixing three components: a uniform distribution  $\mathcal{U}(-40, 40)$ (\Uniform), a positive Gaussian with a mean value 25 and standard deviation of 5, $\mathcal{N}(25,5^2)$ (\GausPos) , and a similar Gaussian with a negative mean, $\mathcal{N}(-25,5^2)$ (\GausNeg). %{\nvk{Is this notation commonly followed for Gaussian distribution?} \ms{added some clarity}} 
We chose a few representative values for the mixing proportions and the total number of sources for the various populations. The boundaries of the uniform distribution or the mean and variance of the two Gaussian distributions do not carry any direct physical significance, rather these are chosen for the sake of diversity in the non-BBH signals.  

All the six populations and their $\dkOne$ and $\dkTwo$ distributions are summarised in  Table~\ref{tab:pop_properties}. Note that we do not impose  $\dkOne = \dkTwo$ (or $\dka=0$) for the simulated signals, to keep them as generic as possible, though our analysis framework makes this assumption. 
In general, the compact binary distribution in the universe could be diverse, characterized by different values of $\dks$  and $\dka$. We create various population models to mimic such scenarios by allowing the fraction of non-BBH systems to differ from model to model.  
{We list all the six population below. You may refer to Table~\ref{tab:pop_properties} for the details of their ingredients.
 
    	\begin{table}
		\begin{tabular}{ccccccc}
			\hline\hline
			\addlinespace[1mm]
			{Model } & \hspace{0.2mm} \textbf{$\fnbh$} \hspace{0.2mm}& \hspace{0.2mm} {BBH} \hspace{0.2mm}& \hspace{0.2mm} {Uniform} \hspace{0.2mm}& \hspace{0.2mm} {GausPos} \hspace{0.2mm}& \hspace{0.2mm} {GausNeg}  \hspace{0.2mm} & \hspace{0.2mm} {$N_{tot}$} \\
			\hline\hline
			\addlinespace[1mm]
			\multirow{1}{*}{\BBH} &0.0 &50 &0  &0  &0  & 50\\ \addlinespace[2mm]\hline\addlinespace[1mm]
			\multirow{1}{*}{\NonBBH} &1.0 &0  &20 &15 &15 & 50\\ \addlinespace[2mm]\hline\addlinespace[1mm]
			\multirow{1}{*}{\NonBBHPos} &1.0 &0  &10 &15 &0  & 25\\ \addlinespace[2mm]\hline\addlinespace[1mm]
			\multirow{1}{*}{\NonBBHNeg} &1.0 &0  &10 &0 &15  & 25\\ \addlinespace[2mm]\hline\addlinespace[1mm]
			\multirow{1}{*}{\MixtureAll} &0.5 &30 &10 &10 &10 & 60\\ \addlinespace[2mm]\hline\addlinespace[1mm]
			\multirow{1}{*}{\MixturePos} &0.5 &20 &10 &10 &0  & 40\\ %\addlinespace[2mm]\hline\addlinespace[1mm]
			\addlinespace[2mm]
			\hline\hline
		\end{tabular}
		\caption{{Details of the simulated compact-binary populations used in this study.  Our six populations are labelled as written in the left-most column. The second column provides the fraction ($\fnbh$) of non-BBH signals contained in each population. The  third column provides the number of BBH signals in each population. The $\dkOne$ and $\dkTwo$ values of the non-BBH signals in each population are distributed as a mixture of three statistical models: a uniform distribution  $\mathcal{U}(-40, 40)$ (\Uniform), a positive Gaussian  $\mathcal{N}(25,5^2)$ (\GausPos), and a negative Gaussian $\mathcal{N}(-25,5^2)$ (\GausNeg). The columns 4-6 describes respective numbers drawn from each of these model. Finally, the  right-most column gives the total size of the population ($N_{tot}$)}. 
		}
		\label{tab:pop_properties}
	\end{table}

\begin{enumerate}
\item 
\BBH: A fully BBH population.
\item 
\NonBBH: A fully non-BBH population with $\dkOne$ and $\dkTwo$ drawn from \Uniform, \GausPos~ and \GausNeg. 
\item 
\NonBBHPos: A fully non-BBH population with  $\dkOne$ and $\dkTwo$ taking only positive values, i.e., from \GausPos~and \Uniform~with a restriction that $\dkOneTwo > 10$.  
\item 
\NonBBHNeg: Afully non-BBH population with  $\dkOne$ and $\dkTwo$ taking only negative values, i.e., from \GausNeg~and \Uniform~with a restriction that $\dkOneTwo < -10$.  
\item 
\MixtureAll: A population containing 50\% BBH signals and 50\% non-BBH signals whose $\dkOne$ and $\dkTwo$ are drawn from \Uniform, \GausPos~and \GausNeg.
\item 
\MixturePos: A population containing 50\% BBH signals and 50\% non-BBH signals whose
$\dkOne$ and $\dkTwo$  are taking only positive values, from  \GausPos, and  \Uniform~ with a restriction that $\dkOneTwo > 10$.

\end{enumerate}

}
We use the \texttt{IMRPhenomPv2} waveform model~ \cite{Ajith:2009bn, Husa:2015iqa, Hannam:2013oca,lalsuite} for simulating all the signals, which is the same as the model used for the Bayesian analysis as well, as mentioned before. 
\begin{figure*}
    \centering
    \includegraphics[scale=0.5]{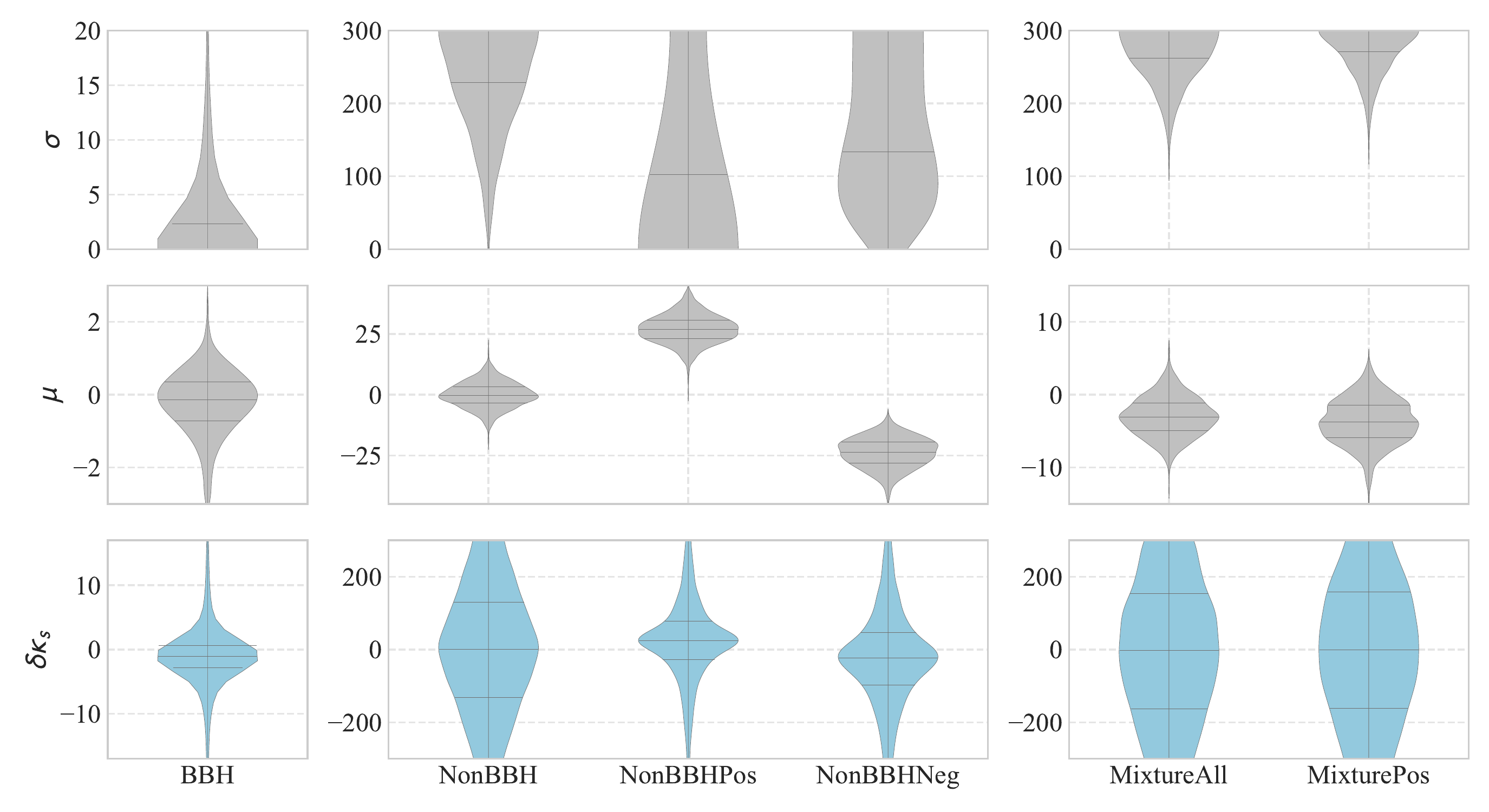}
    \caption{Violin plots showing the results from hierarchical combining formalism applied on the six simulated populations as labelled on the x-axis. The top two rows (silver) are the posterior densities of the hyper parameters $\mu$  and $\sigma$ characterising the \emph{Gaussian} that models the population distribution  $\dks$ (see  Sec.~\ref{sec:method-hier} and Eq.~(\ref{eq:gaussian-single}) for details). 
    The bottom row (blue) shows the $\dks$ distributions obtained by marginalizing over the  $\mu$ and $\sigma$ posteriors (as per Eq.~(\ref{eq:post-dks})). We assumed $\mu$ and $\sigma$ to have uniform priors in the ranges $[-150,150]$ and $[0,300]$  respectively and sampled using the \texttt{Dynesty} sampler from \texttt{Bilby}
}
\label{fig:hiercombining}
\end{figure*}

\section{Results}
\label{sec:results}

\subsection{Hierarchical combining approach}
We apply the method described in section~\ref{sec:method-hier} on all the six simulated populations described in Table~\ref{tab:pop_properties}. The results are shown as violin plots in Figure~\ref{fig:hiercombining}. The top two rows (silver) show the posterior distributions of the hyper-parameters $\mu$ and $\sigma$ for each population and the bottom panel shows the  $\dks$ distributions reconstructed from $\mu$ and $\sigma$ according to Eq.~(\ref{eq:post-dks}).
For \BBH, the distributions of both $\mu$ and $\sigma$ peak at their true values (i.e., zero) with narrow error bars and so is the  $\dks$ distribution which is  constrained to [-10.7, 8.8] at 90\% credibility. However, this is not the case with the rest of the populations, as we discuss  below. 

For the \NonBBH, the $\mu$ posterior is consistent with zero which is expected because it has equal number of sources from both sides of $\dks=0$. The $\sigma$ posterior has increasing support towards the higher end of the prior and there is little to no support for $\sigma=0$. This implies that the population could not be fit with a simple Gaussian and can be taken as an indication to the presence of one or more of sub-population with different values of $\dks$. The marginalized $\dks$ distribution for this case is a wide distribution with no insight, which results from the behaviour of the $\sigma$ posterior. 

For the  \NonBBHPos~and  \NonBBHNeg, the $\mu$ tends to peak at their true values again but the $\sigma$ posteriors do not exclude zero. This is because the injected distributions of $\dks$ are one-sided for these two  ($\sim[0,40]$ and $\sim[-40,0]$) unlike \NonBBH~  ($\sim[-40,40]$)  and hence are relatively better candidates for finite-width Gaussians. Furthermore, we see that the peaks of the $\dks$ distributions are shifted to positive and negative values respectively for \NonBBHPos~ and \NonBBHNeg, as one would expect, though they do not exclude zero. 

For the mixture populations (\MixtureAll~and \MixturePos), the $\sigma$ posteriors completely excludes zero and peaks at the highest value allowed by the prior. This again shows that a Gaussian of finite width could not be used to represent the underlying $\dks$ distribution and hence one could conclude the population to have complex components. The reconstructed $\dks$ distributions are, as expected, uninformative.   

In summary, the \textit{hierarchical combining approach}  provides reasonably good estimates when we have only BBH signals or only non-BBH signals with all of them belonging to similar nature, i.e., $\dks$ having only positive/negative values in a small range of values like the ones we considered. For mixture population containing both BBH and non-BBH signals, the method can indicate to the complexity of the underlying distributions however can not realize whether it is full of non-BBH or a mixture of both BBH and non-BBH.

\begin{figure}
    \centering
    \includegraphics[scale=0.55]{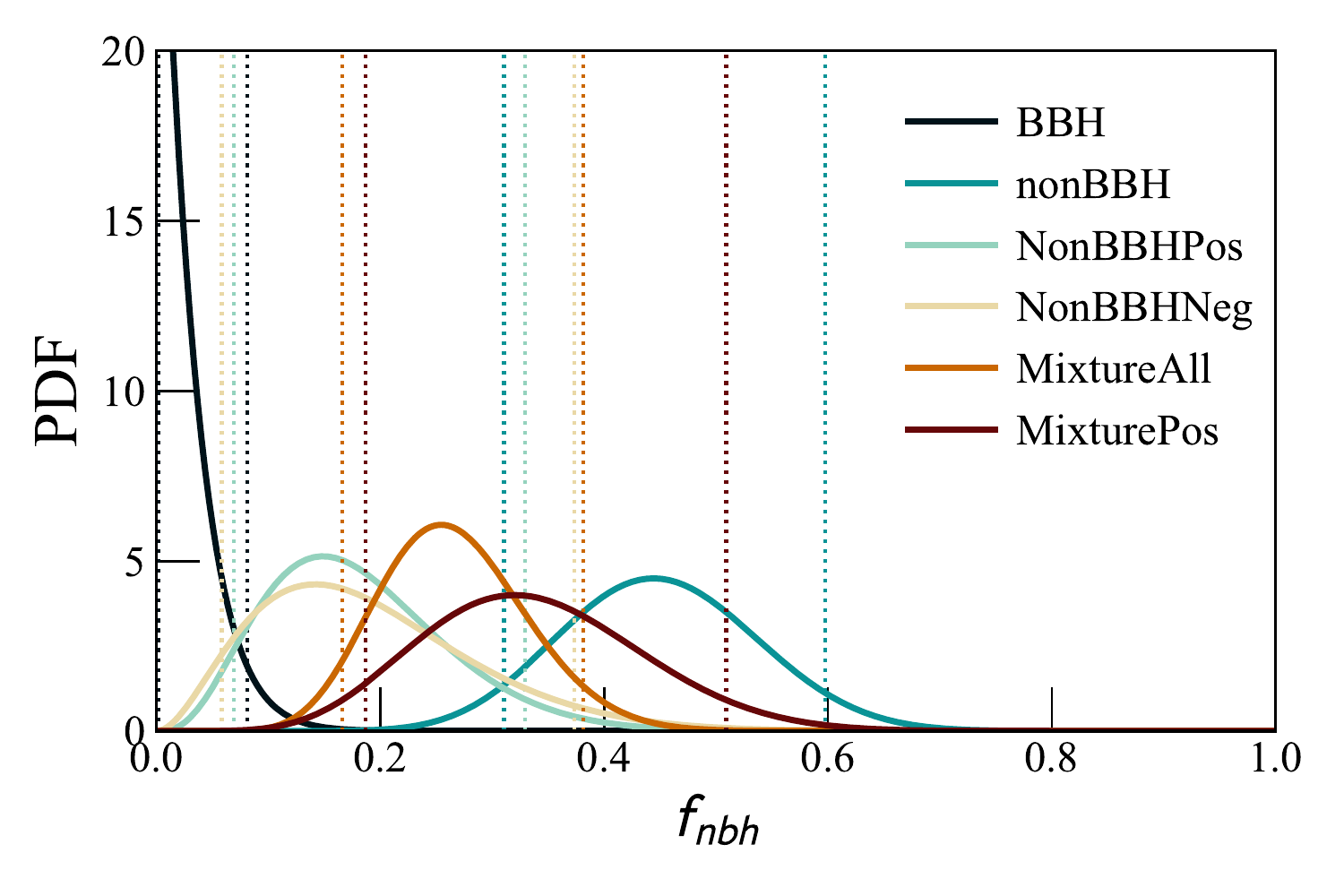}
    \caption{{Posterior distributions of the $\fnbh$ parameter - the fraction of non-BBH signals in the population - estimated using the \emph{mixture-likelihood} approach. For a population with only BBH signals, the posterior is expected to peak at zero. A peak at unity would imply to a population with only non-BBH signals while a peak in between indicating a mixture population with BBH and non-BBH signals. The posteriors shown here are for the six simulated compact binary populations  described in Table~\ref{tab:pop_properties}. The vertical dashed lines represent the  90\% credible intervals.}}
\label{fig:fbh}
\end{figure}

\subsection{Mixture likelihood approach}
Now, we apply the \emph{Mixture likelihood} method ( section~\ref{sec:method-fraction}) on all the six simulated populations of Table~\ref{tab:pop_properties}. Fig.~\ref{fig:fbh} shows the posteriors on $\fnbh$, the fraction of non-BBH events in the population. We see that, for the  BBH-only population (\BBH), the $\fnbh$ distribution peaks at zero, as expected. For the non-BBH populations with no BBH sources at all (\NonBBH, \NonBBHPos~and \NonBBHNeg), we expect $\fnbh$ to peak at unity. However, the peaks occur between 0.1 and 0.5. Similarly, for the mixture populations with an equal number of BBH and non-BBH sources (\MixtureAll~and \MixturePos), the peaks are expected to be at 0.5, but we obtain the peaks between 0.2 and 0.4. This shows that there is an overall tendency for $\fnbh$ to lean towards the BBH value. These offsets can be understood as systematics due to multiple reasons and have been discussed in detail in Sec.~\ref{sec:bias}. 
Nevertheless, for all the non-BBH populations we have chosen, the $\fnbh$ posteriors exclude zero at 90\% credibility.

\begin{figure}
    \centering
    \includegraphics[scale=0.55]{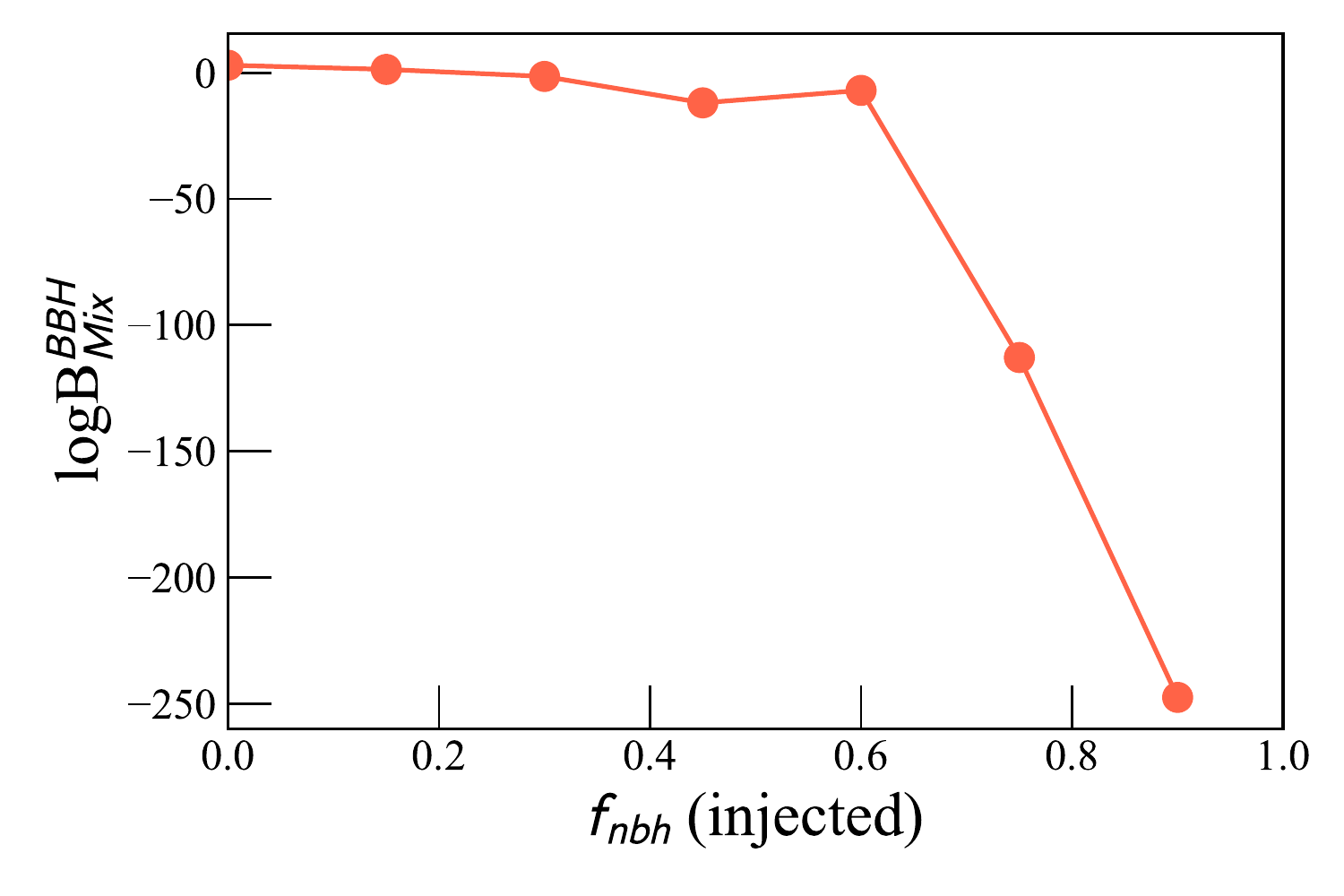}
    \caption{
    Figure showing how the log$_e$ Bayes factor between the population-hypotheses ``All are BBH'' and  ``A mixture of BBH and non-BBH'' (as derived in Eq.~(\ref{eq:logB-bh-vs-mix})) varies as a function of the true value of the $\fnbh$ of the population. The log$_e$ Bayes factors shown on the y-axis, at each $\fnbh$ on the x-axis, are ensemble-averaged over 500 population instances, where each instance has $N_{tot} = 30$ sources. The 500 instances are constructed by collecting the simulated events across the six populations of Table~\ref{tab:pop_properties}. Of course, the ensembles are not mutually exclusive as we have a limited number of sources in the population. Nevertheless, the averaging helps reduce the outlier effects, which would be expected with a population size as small as 30.}
    %\abhi{What does this mean: "500 random mixtures such that the outliers are averaged out"? The plot label shows Ntotal=30. Does that mean you use 30 events in this population? Does it correspond to a particular PopX from the table? How do you do 500 random mixture of a sample of 30 events? And what are the outliers?}. \ms{Hope the updated caption and text clarifies these points} }
    % One can also have the  error bar on logB from the 500 random trials 
\label{fig:bf_fraction_plot}
\end{figure}

In Fig.~\ref{fig:bf_fraction_plot}, we show the log$_e$ Bayes factor between the ``All BBH'' vs the ``Mixture of BBH and non-BBH'' hypotheses as derived in Eq.~(\ref{eq:logB-bh-vs-mix}),  as a function of the injected value of $\fnbh$. Note that we do not use any of the populations in Table~\ref{tab:pop_properties}  for this plot. Rather we follow an averaging procedure using the sources in all those populations. We first construct two pools of BBH and non-BBH sources by collecting all available sources from the six populations in Table~\ref{tab:pop_properties}. Now, for a given value of $\fnbh$ on the x-axis, for example, $\fnbh=0.4$, we randomly pick 12 BBH and 18 non-BBH sources from the respective pools, with a total of 30 sources. By repeating this 500 times, we compute log$_e$ Bayes factor for all these 500 cases and then take the average of them, and is shown on the y-axis of Fig.~\ref{fig:bf_fraction_plot}. The averaging helps remove the fluctuations due to any outliers when considering a population size as small as 30.
%\ms{Not sure we should show the error bars?}   

The log$_e$ Bayes factor favors the ``only BBH'' hypothesis when we inject only BBH signals (at $\fnbh=0$), whereas it rules out the ``only BBH" model for all the populations that included non-BBH signals (as we move towards larger values of $\fnbh$).

\begin{figure}
    \centering
    \includegraphics[scale=0.55]{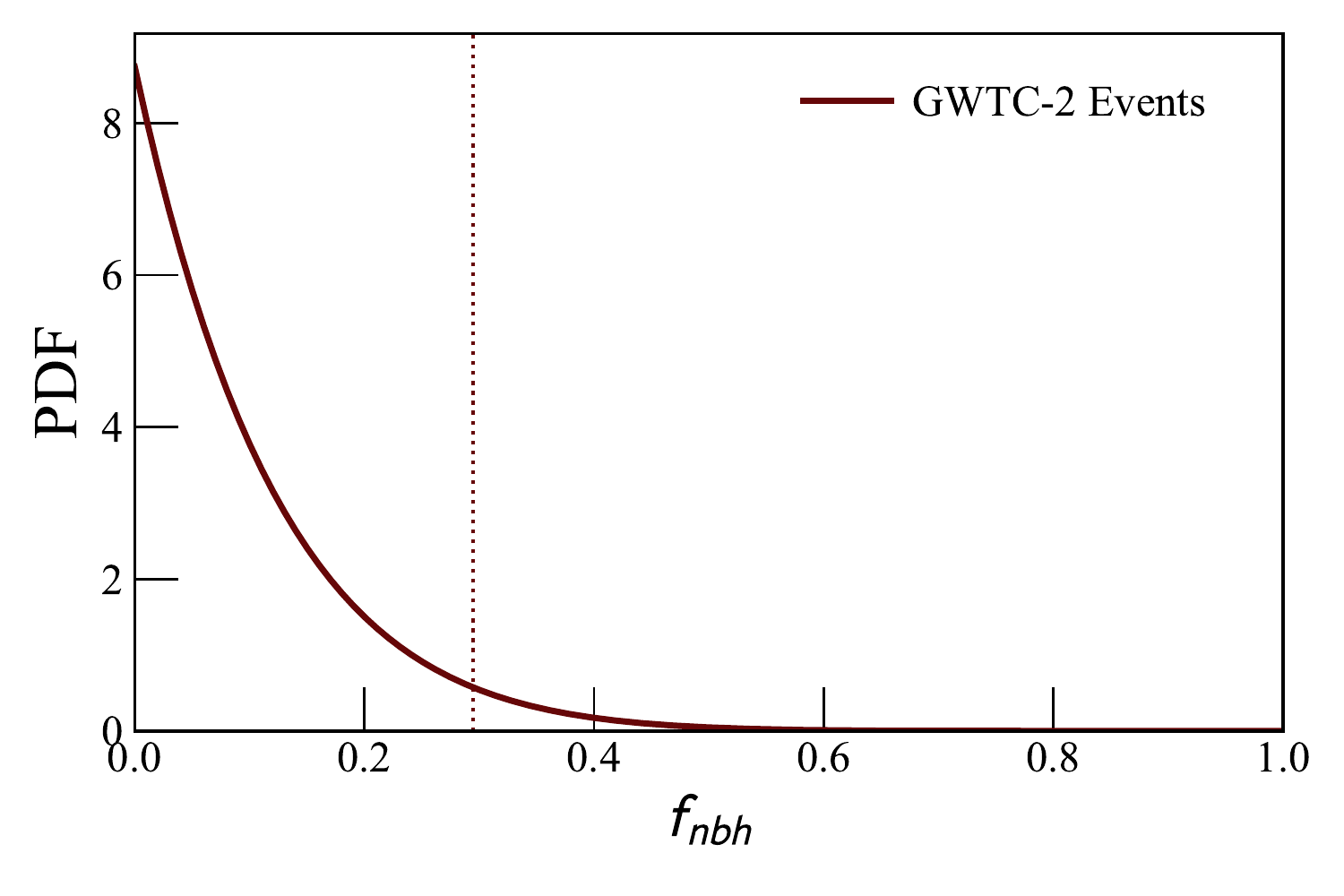}
    \caption{Same as Fig.~\ref{fig:fbh} but the posterior of $\fnbh$ for the compact binary detections of Advanced LIGO and Advanced Virgo reported in the {GWTC-2 considered for testing GR analysis}~\cite{LIGOScientific:2020tif}.  The peak of the posterior, as obtained here, is consistent with a population of only BBH signals.}
\label{fig:fbh_realevent}
\end{figure}

\begin{figure*}
    %\centering
    \includegraphics[scale=0.56]{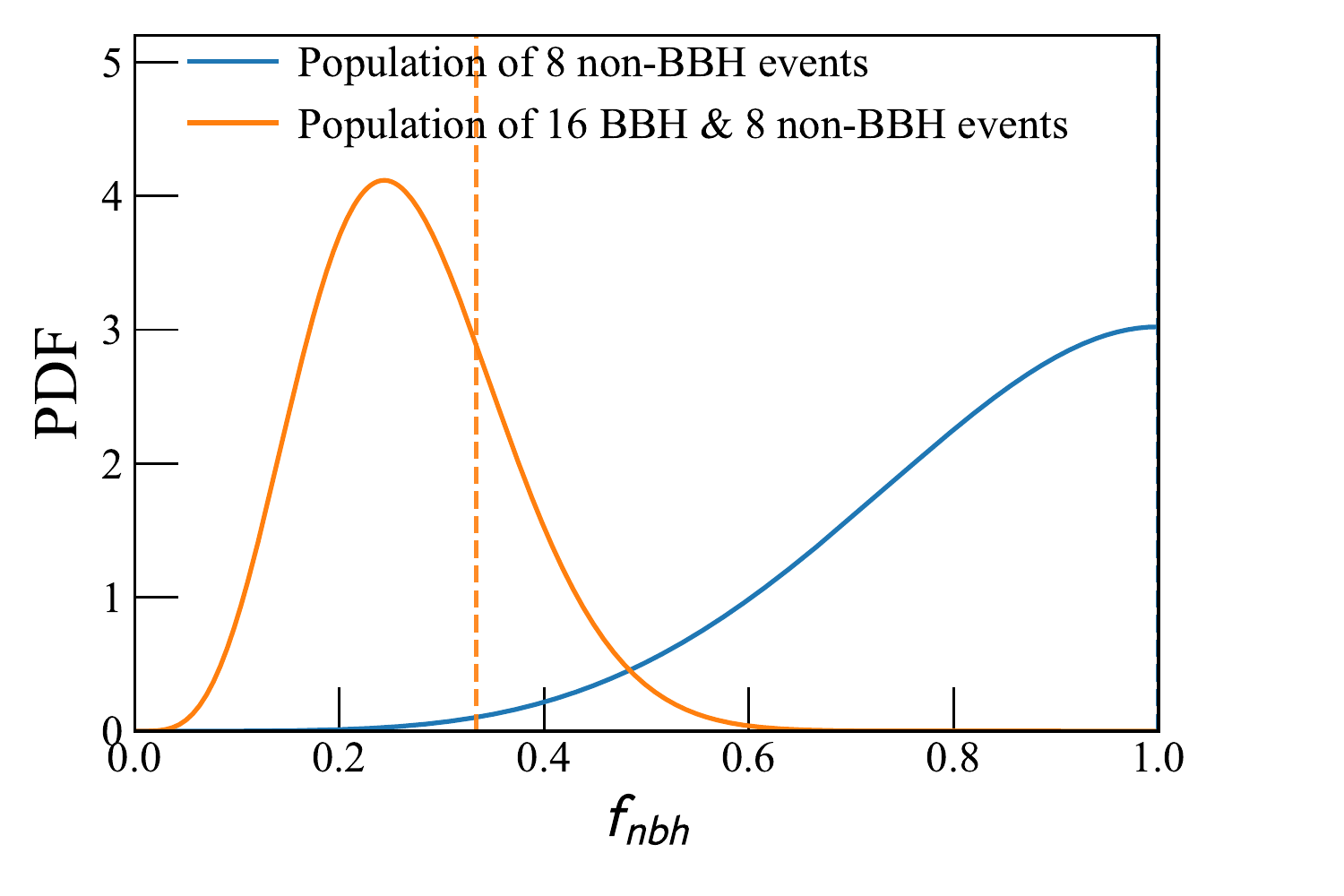}
    \includegraphics[scale=0.56]{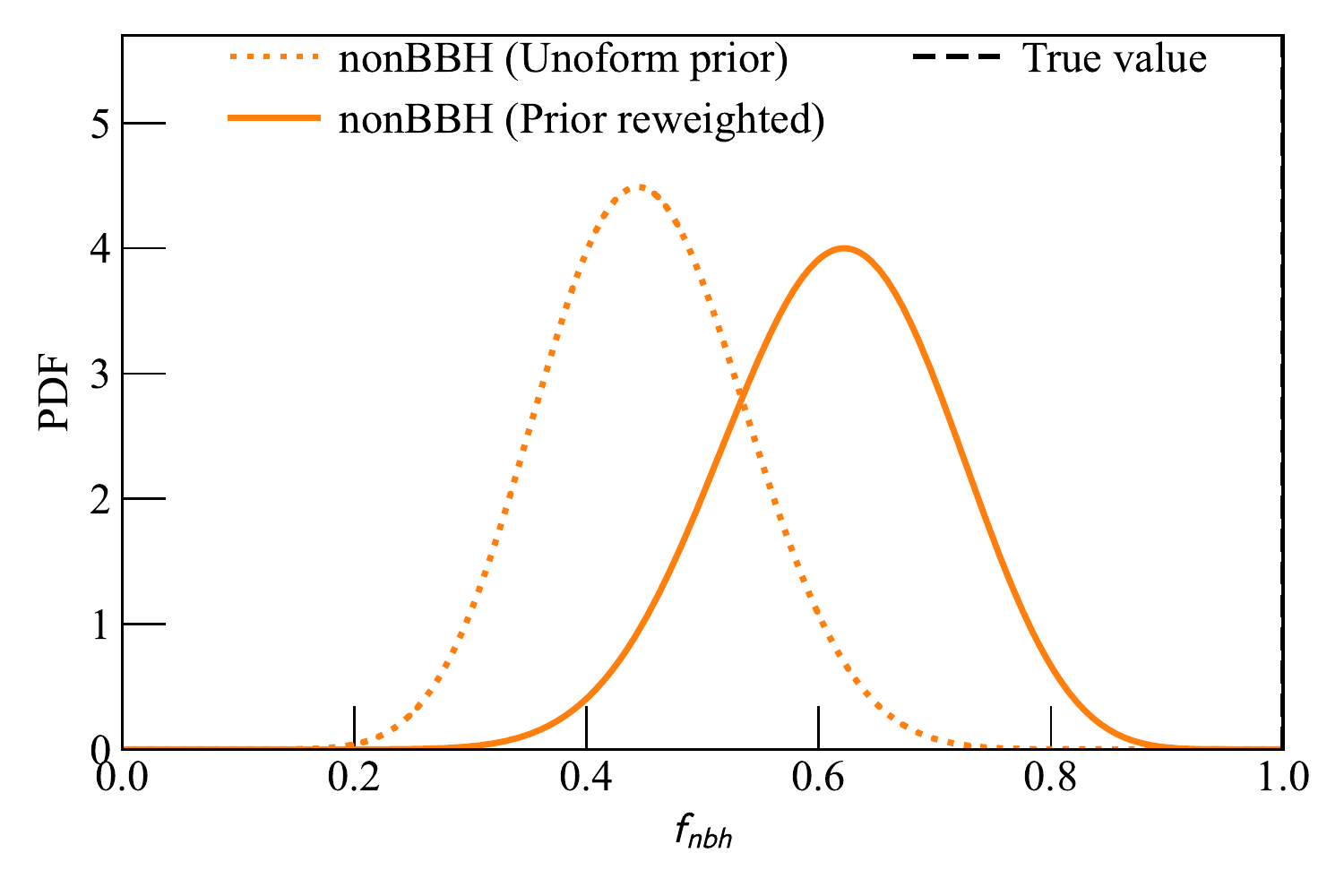}
    \caption{[Left] 
    %\ms{The figures need to be made with same aspect ratios}
    {Posterior distribution of $\fnbh$ for populations that include simulated sources with $\vert \chi_{\rm{eff}}\vert \geq 0.15$ and  network $\rm{SNR_{\rm{insp}}}\geq 20$. These populations are constructed by collecting sources from all the six populations that satisfy the SNR and spin criterion.} [Right] The posterior distribution on the fraction $\fnbh$ with and without prior-reweighting discussed in Sec.~\ref{sec:bias} and Appendix ~\ref{App:prioreffect}. On both the panels the dashed vertical lines indicate the injected value of $\fnbh$.  }
\label{fig:systematics}
\end{figure*}

\subsection{Estimation of fraction of non-BBH signals from real LIGO-Virgo observations}

The first bounds on spin-induced quadrupole moment parameter from observed gravitational wave events are reported in ~\cite{Krishnendu:2019tjp}. Furthermore, Ref.~\cite{LIGOScientific:2020tif} provided the combined posterior distributions on $\dks$ obtained from the O1, O2, and the first half of O3 observing runs of LIGO-Virgo detectors along with the individual bounds. In ~\cite{LIGOScientific:2020tif}, the combined bounds and Bayes factors are calculated following two methods: multiplying likelihoods (universality assumption on $\dks$) and hierarchical combining. Here we demonstrate the applicability of the mixture likelihood approach on all the events considered in~\cite{LIGOScientific:2020tif}  and  infer $\fnbh$ for this population. {We use the public data available from ~\cite{GWOSC:GWTC-2}}. The result is shown in Fig.~\ref{fig:fbh_realevent}. Our analysis confirms that the detections reported in GWTC-2~\cite{Abbott:2020jks} are consistent with a BBH population.

\section{Understanding the systematic biases}\label{sec:bias}
The bias in the $\fnbh$ posterior towards the BBH value ($\fnbh=0$)  as discussed above and shown in Fig.~\ref{fig:fbh} could be a sum of various effects in play.  We discuss these effects in the present section.

\subsection{Biased single-event inferences}
 
     %Fig.~\ref{fig:systematics}
One of the possible reasons for the systematic bias in the inference of $\fnbh$ is the choices of spins and SNRs of the individual events in the populations. It has been shown that the single-event $\dks$ inferences are better when the spins are higher, and of course, improves further with higher SNR~\cite{Krishnendu:2019ebd} while our populations include many events with low SNRs and low spins.
     
To investigate this, we collect events from our simulations that have effective spins $\vert \chi_{\rm{eff}}\vert \geq 0.15$ and the network $\rm{SNR_{insp}} \geq 20$~{\footnote{$\chi_{\rm{eff}}$ is the effective inspiral spin parameter captures the spin effects of non-precessing binary system which is defined in terms of component masses $m_{i}$ and dimensionless spins $\chi_{i}=\vec{S_{i}} \cdot \hat{L} / m_{i}^2$ as, $\chi_{\rm{eff}} = \frac{m_{1}\chi_{1}+m_{2}\chi_{2}}{m_1+m_2}$~\cite{Ajith:2009bn}.}}. Out of all the BBH and non-BBH simulations that span over six populations, we obtain 16 BBH and 8 non-BBH events satisfying the above criterion. We construct two populations out of them: (1) All the 8 non-BBH events together, with a true value of $\fnbh=1$, and (2) all the 16 BBH and 8 non-BBH events together, with $\fnbh=0.33$.
     
We estimate $\fnbh$ for these two populations using the mixture-likelihood approach and find that the posteriors are consistent with the true values at very high confidences, as illustrated on the left panel of Fig.~\ref{fig:systematics}. The fact that we only had 8 and 24 events for these two populations has been reflective in the respective statistical uncertainties in  Fig.~\ref{fig:systematics} (left panel). This finding indicates that the bias in $\fnbh$ primarily arises because our population contains many low-spin-low-SNR events.

\subsection{Prior effects} 
Throughout the Bayesian inference, we have assumed the prior on $\dks$ as uniform in [-500,500]. The injected populations, on the other hand,  have $\dks$ drawn from narrower ranges with non-uniform distributions. Prior ranges that are wider than required will be penalized by Ocaam's factor. Below, we show that our prior choices can partly account for the bias in $\fnbh$ estimates. The right panel of Fig.~\ref{fig:systematics} shows how the $\fnbh$ posterior would change if we were to perform the analysis with the same prior as that of the injected $\dks$ distribution. In order to achieve this, we did a prior re-weighting as described in Appendix~\ref{App:prioreffect}. We have taken an example from our populations with $\fnbh=1$ and find that the posterior on $\fnbh$ after prior re-weighting (dashed curves) is closer to the true values.

Of course, in reality, we can not pre-acquire the knowledge of the underlying distribution. However, we can always assume a hyper-parametrized prior model characterizing the underlying distribution and marginalize the likelihood over the hyperparameters  to obtain the posteriors on $\fnbh$. In Appendix~\ref{App:prioreffect}, we have derived the formalism of performing this hyperparameter-marginalization though its detailed demonstration is differed for a future work.

	\subsection{ Non-identical nature of the binary components}
	Another possible, though minor,  reason is our assumption of $\dka=0$ (i.e., $\dkOne = \dkTwo$) in the analysis while our injections did not assume this. In other words, we have injected different values for $\dkOne$ and $\dkTwo$ while in the Bayesian sampling, we assumed them to have equal values. This can lead to a bias in the estimated $\dks$ and the Bayes factor. Indeed, this is a prospect to explore in detail in a future study. Regardless of the $\dka=0$ assumption, the posterior could capture the true value of $\fnbh$, if the SNR and spins are high enough, as shown in the left panel of Fig.~\ref{fig:systematics}.

At higher detector sensitivity and with better waveform models such as those with higher modes~\cite{Shaik:2019dym,Varma:2016dnf,BustilloHusaPurrer_etal_HM_alignedspin:2015qty,CollinCapano_HMs_NoSpin:2013raa,Littenberg:2012uj,Kalaghatgi:2019log}, some further correlations  between binary parameters are expected to break, leading to improved estimates of $\kappa$ parameters and improved Bayes factors, for the individual events. This will eventually improve the measurement of $\fnbh$.

\section{Conclusion and further remarks}
\label{sec:conc}
We have discussed how to effectively \emph{combine} gravitational-wave data from multiple detections to probe sub-populations that include non-BH compact binaries. Our method is based on using spin-induced quadrupole moment as a physical parameter that distinguishes BBH from non-BBHs. We first analyzed the efficacy of the previously-employed hierarchical combining approach that relies on a Gaussian assumption for the population distribution of the spin-induced quadrupole moment parameter. Next, we introduced a mixture-likelihood approach (Sec.~\ref{sec:method-fraction}) that estimates the fraction ($\fnbh$) of non-BBH signals present in the observed population. 

We simulated various populations that included BBH and non-BBH signals at different proportions. Our results show that both approaches are good for homogeneous populations like a BBH-only population. Also, both the approaches would signal if the population has complex nature with subpopulations being present. The mixture-likelihood approach is a natural choice to capture such complex distributions of non-BBH sub-populations. We applied the method on the LIGO-Virgo detected GW events from the GWTC-2 catalog and found them consistent with a BBH population.

Though the mixture-likelihood approach effectively rules out the BBH population hypothesis for the simulated populations that included some fraction of non-BBH signals, we notice that the method suffers from systematics in measuring the fraction of non-BBH signals precisely.  We investigated possible reasons for these biases in Sec.~\ref{sec:bias}.

In the future, with more realistic astrophysical population models, our estimates may improve. This is anticipated primarily because the spin-induced quadrupole moment estimates from individual events and, hence, the population can change according to the intrinsic mass-spin distributions. Also, with more accurate waveform models being available, our measurements could further improve as they could help break some of the degeneracies that lead to the systematics in this study.  We would also include the effect of selection bias in the inference of $\fnbh$ which would allow re-defining the \emph{fraction} $\fnbh$ as the fraction of non-BBH signals in the universe, rather than the fraction in the \emph{observed} signals, in the future. 

Finally, though we have demonstrated the mixture-likelihood approach using the measurement of spin-induced quadrupole moment parameters, the method is generic enough to include in or to be applied for the other physical parametrizations that distinguish BBH from non-BBH signals (as mentioned in Sec.~\ref{intro}).  

\acknowledgements

	The authors are grateful to Frank Ohme for useful comments on the manuscript. 
The authors acknowledge the use of LIGO LDG clusters for the computational work done for this study. This research has made use of data obtained from the Gravitational Wave Open Science Center (www.gw- openscience.org), a service of LIGO Laboratory, the LIGO Scientific Collaboration and the Virgo Collaboration. LIGO is funded by the US NSF. Virgo is funded by the French Centre National de Recherche Scientifique (CNRS), the Italian Istituto Nazionale della Fisica Nucleare (INFN) and the Dutch Nikhef, with contributions by Polish and Hungarian institutes. 	
	
	M. Saleem acknowledges support from NSF grants PHY-00090754, PHY-1806630, and PHY-2010970, and the support from the Infosys Foundation, the Swarnajayanti fellowship grant DST/SJF/PSA- 01/2017-18. 
	N. V. Krishnendu acknowledges support from the Max Planck Society’s Independent Research Group Grant.
	The research of Archisman Ghosh~is supported by Ghent University's BOF Starting Grant BOF/STA/202009/040.
	K. G. Arun partially supported by a
	grant from Infosys Foundation. K. G. Arun also acknowledges the
	Swarnajayanti fellowship grant 
	DST/SJF/PSA-01/2017-18, and 
	EMR/2016/005594  and MATRICS grant
MTR/2020/000177 of SERB.

This document has LIGO preprint number {\tt LIGO-P2100382}.

\appendix

\section{Marginalizing the mixture likelihood over the hyper parameters of $\dks$ distribution}
\label{App:prioreffect}

In Eq.~(\ref{eq:fnbh_post}), we have considered a uniform prior on $\dks$ in estimating the evidence $\Znbh $. In a more generic treatment, we should hyper-parametrize this prior and marginalize over, as the underlying $\dks$ distribution in the universe is unknown (See Ref.~\cite{Smith:2020lkj} for a comprehensive treatment of marginalizing over the prior hyper-parameters, though in a different context). Thus, we first rewrite Eq.~(\ref{eq:fnbh_post}) with a generic prior on $\dks$, labelled as $\Lambda$. That means, Eq.~(\ref{eq:likelihood-mixture}) takes the form,

\begin{equation}
\mathcal{L}^{\text{pop}}(\dataN \vert \fnbh, \Lambda) = \prod_{i=1}^N \left[ (1-\fnbh) \, \mathcal{Z}_i^{bh} + \fnbh \,\, \mathcal{Z}_i^{nbh} (\Lambda)\right],
\label{eq:likelihood-mixture-Lambda}
\end{equation}
where $\mathcal{Z}_i^{nbh} (\Lambda)$ is the generalized version of Eq.~(\ref{eq:evidence}),
\begin{equation}
        \Znbh (\Lambda)= \int \pi(\pnbh|\Lambda) \, \mathcal{L}(d \vert \pnbh)\, d\pbh\, d\delta\kappa_s.
        \label{eq:evidence-Lambda}
    \end{equation}

The only difference being that the prior on $\dks$ is now informed by the hyper parameters $\Lambda$. Also, this hyper parameter space $\Lambda$ could in general be a complex multi-dimensional parameter space. Perhaps, the simplest case one could consider is a zero-centered top-hat function, with only one parameter, namely $\lambda$.  This would be a prior pretty much like the one we have used in our analysis, $\pi(\dks|\lambda)\sim U(-\lambda, \lambda)$. Alternatively, if one assumes that all the non-BBH signals are from a certain class of exotic compact objects and have their $\dks$ distribution coming from a localized distribution, then a Gaussian with unknown mean and width would be a good representation, \textit{i.e.},
$\Lambda=\{\mu, \sigma^2\}$ 
%\ag{$\sigma^2$?}.
Given a hyper-parameter model, the corresponding evidence $\Znbh (\Lambda)$ at any given point in the $\Lambda$-space can be evaluated by re-weighting the prior, as discussed in Appendix~\ref{sec:reweight}). 

Once we have the tools to evaluate the likelihood, we can sample from the likelihood using any sampler, over the parameters, $\{\fnbh, \Lambda\}$. This would give us the posterior  probability density function as, 
    \begin{equation}
        p(\fnbh, \Lambda|{\dj}) = \int \frac{\mathcal{L}^{pop}(\dj|\fnbh,\Lambda)\, \pi(\fnbh)\, \pi(\Lambda)}{\mathcal{Z}^{pop}_{\Lambda}}, 
        \label{eq:fnbhposteriorNewLambda}
    \end{equation}
    with 
    %\sout{$\mathcal{Z}^{pop}_{\Lambda}$},
    \begin{equation}
        \mathcal{Z}^{pop}_{\Lambda} = \int \mathcal{L}^{pop}(\dj|\fnbh,\Lambda) \, \pi(\fnbh)\, \pi(\Lambda) \,d\fnbh \,d\Lambda.
    \end{equation}
    Now we can obtain the posterior on $\fnbh$ marginalized over $\Lambda$ as,
    \begin{equation}\label{eq:fnbhLambda}
        p(\fnbh|\dj) = \int p(\fnbh, \Lambda| \dj) d\Lambda\,.
    \end{equation}
    Employing a nested sampling algorithm for the sampling, from the likelihood given in Eq.~(\ref{eq:evidence-Lambda}) the output will by default provide the probability of fraction defined in Eq.~(\ref{eq:fnbhposteriorNewLambda}). The priors on $\fnbh$ and $\Lambda$ can be taken as uniform. Given the posterior distribution on $\dks$, we can also reconstruct the population distribution of $\dks$ as, 
    \begin{equation}
        p(\dks|\dj)=\int p(\dks|\Lambda) \, p(\Lambda|\dj) \, d\Lambda\,.
        \label{eq:dksposterior_Lambda}
    \end{equation}
    Equation~(\ref{eq:dksposterior_Lambda}) is identical to Eq.~(\ref{eq:post-dks}) with the only visible difference being the hyper parameter vector $\vec{\alpha}$ replaced by $\Lambda$. The different notions of hyper parameters in Eq.~(\ref{eq:dksposterior_Lambda}) and Eq.~(\ref{eq:post-dks}) are purposeful, while $\vec{\alpha}$ represents the distribution of $\dks$ from the entire population, $\Lambda$ by definition represents the distribution of $\fnbh$, the fraction of non-BBHs in the population. The remaining $(1-\fnbh)$ fraction of sources have a delta function prior, given by $\delta(\dks-0)$.

\subsection{Evidence estimation by prior re-weighting}\label{sec:reweight}

Suppose we have the evidence estimated already assuming one prior, in our case a prior on $\dks$ assuming U[-500, 500]. Let us call this prior as $ \Phi$. Now the evidence for this prior $\Znbh(\Phi)$, according to the Bayes theorem, can be written as, 
\begin{equation}
         \Znbh (\Phi) \, p( \pnbh\, | \dj,\, \Phi) = \pi( \pnbh\, | \Phi ) \, \mathcal{L}(\dj\, |\, \pnbh),
        \label{eq:evidence-phi}
    \end{equation}
With the prior $\Lambda$, the new evidence will be as given by Eq.~(\ref{eq:evidence-Lambda}). Plugging Eq.~(\ref{eq:evidence-phi}) to Eq.~(\ref{eq:evidence-Lambda}), we get

\begin{equation}
    \Znbh (\Lambda)= \Znbh (\Phi)\int  \frac{\pi(\pnbh|\Lambda) }{\pi( \pnbh\, | \Phi )} \, p( \pnbh\, | \dj,\, \Phi) d\pbh\, d\delta\kappa_s \,,
    % \Znbh (\varphi)\int \frac{p( \pnbh\, | \dj,\, \varphi)}{\pi( \pnbh\, | \varphi )} \,\pi(\pnbh|\Lambda)\, d\pbh\, d\delta\kappa_s.  \nonumber 
\end{equation}
where the probability distribution $p( \pnbh\, | \dj,\, \Phi) $ is known and the above integral can be approximated as a Monte-Carlo average over the posterior samples, \textit{i.e.},
\begin{equation}
    \Znbh (\Lambda)= \Znbh (\Phi) \, \times \left[ \frac{1}{n} \, \sum_{k=1}^{n} \frac{\pi(\pnbh|\Lambda)}{\pi(\pnbh|\Phi)}\right],
    \label{eq:evidence_lambda_phi}
\end{equation}
where $n$ is the number of samples in the $\dks$ posterior distribution. Both $\pi(\pnbh|\Lambda)$ and $\pi(\pnbh|\Phi)$ must be normalized distributions.
Throughout the analysis, we assume that the priors on $\dks$ and $\vec{\theta}$ are uncorrelated and hence we can decouple  
$\pi(\pbh)$ from both numerator and denominator
so that they cancel each other. This leaves,
\begin{equation}
    \Znbh (\Lambda)= \Znbh (\Phi) \, \times  \left[ \frac{1}{n} \, \sum_{k=1}^{n} \frac{\pi(\dks|\Lambda)}{\pi(\dks|\Phi)}\right],
    \label{eq:evidence_lambda_phi}
\end{equation}
 
 Equation~(\ref{eq:evidence_lambda_phi}) makes it easier to compute the likelihood in Eq.~(\ref{eq:likelihood-mixture-Lambda}) for any instance of $\Lambda$. The reader may refer to Ref.  ~\cite{2019PASA...36...10T} for a detailed treatment of the prior-re-weighting procedure.

\bibliographystyle{apsrev}
\bibliography{main}

\end{document}